# Light tunneling anomaly in interlaced metallic wire meshes

Hafssaa Latioui[(1)], Mário G. Silveirinha[(1,2)*]

[(1)] *Department of Electrical Engineering, University of Coimbra, and Instituto de Telecomunicações, 3030-290 Coimbra, Portugal*

[(2)]*University of Lisbon–Instituto Superior Técnico - Avenida Rovisco Pais, 1, 1049-001 Lisboa, Portugal*

**Abstract**

For long wavelengths three-dimensional connected metallic wire meshes are impenetrable by light and have an electromagnetic response similar to that of an electron gas below the plasma frequency. Surprisingly, here it is shown that when two opaque metallic meshes are spatially-interlaced the combined structure enables an anomalous light tunneling in the long wavelength regime. The effect is due to the destructive interference of the waves scattered by the two wire meshes, which leads to a Fano-type resonance.

---

[*] To whom correspondence should be addressed: E-mail: *mario.silveirinha@co.it.pt*

Metallic grids block the propagation of light when the electric field is aligned with the grid [1-6]. This property is explored in many applications, such as in light polarizers and frequency selective surfaces. For example, a regular array of thin parallel metallic wires is typically used as a linear polarizer since only the light oscillations perpendicular to the wires can go through the structure.

When the wire grid spans the three directions of space – the so-called three-dimensional (3D) connected wire medium [7-11] – the corresponding photonic crystal does not support light states in the long wavelength regime. Indeed, a 3D connected wire metamaterial is completely opaque to light propagation and has an electromagnetic response analogous to that of a free-electron gas [9, 10]. Hence, intuitively one may expect that when two 3D wire meshes are spatially interlaced (Fig. 1a) the full structure should be a better reflector and should scatter light more strongly than the individual components. In fact, common sense suggests that the effects of the individual wire meshes should be "additive" and that the interlaced wire structure will, accordingly, block the light propagation more effectively. Surprisingly, here we theoretically demonstrate that even though each individual wire mesh exhibits an almost zero transmission for low frequencies, the interlaced meshes can be nearly transparent to electromagnetic waves.

To unveil the physical mechanisms that enable the anomalous light tunneling, first we computed the photonic band diagram of the interlaced wire meshes (Fig. 1b) using the eigenmode solver of CST Microwave Studio [14]. For simplicity, we only provide results for the case where the wave vector is directed along the $\Gamma X$ direction. Each wire mesh is formed by a 3D array of connected metallic wires with lattice constant $a$. The two wire meshes ($A$ and $B$) are disconnected and separated by a distance of $a/2$ along the directions parallel to the coordinate axes (see the inset of Fig. 1b). For now, the wires are assumed perfect electric conductors (PEC). The host dielectric is air ($\varepsilon_h = 1$) and the radii of the wires is $r_{w,A} = 0.001a$ for mesh A and $r_{w,B} = 0.05a$ for mesh B.

Strikingly, Fig. 1b shows that the interlaced wire medium supports an electromagnetic mode for arbitrarily low frequencies. In contrast, it is shown in the supplementary materials [15] that the individual meshes $A$ and $B$ do not support light states in the long wavelength regime. The emergence of low frequency modes in metallic wire arrays with disconnected components was noticed in several previous works [9, 13, 16-19], and in particular Ref. [13] demonstrated that the number of modes supported by a generic 3D wire mesh with $N$ metallic components is identical to $N-1$. Importantly, as shown in Fig. 1c*i*, the low-frequency mode supported by the interlaced wire medium is longitudinal such that on average the electric field has the same orientation as the wave vector **k** [13]. The interlaced wire medium also supports three high frequency modes, specifically two degenerate transverse electromagnetic (TEM) modes and an additional longitudinal mode. The field profiles of these modes are shown in Figs. 1c*iii* and 1c*ii*, respectively. The high frequency modes only propagate above an effective plasma frequency given by $\omega_{p,ef} \approx 2.3c/a$, and their dispersion and polarization is consistent with what is expected of a (spatially-dispersive) 3D electron gas [9].

Clearly, the coupling between the two wire meshes leads to the emergence of a longitudinal propagating state in the low frequency limit. Can however this mode be excited by an incoming plane wave propagating in air? To answer this question, we used CST Microwave Studio to find the reflection and transmission coefficients for an incident plane wave with magnetic field directed along *x* (transverse magnetic – TM – polarization) as a function of the incidence angle. The plane of incidence is the *yoz* plane and the geometry of scattering problem is sketched in the inset of Fig. 2a. The normalized frequency is $\omega a/c = 1.32$ and the thickness of the interlaced wire medium slab is $L \approx 6a$. Counterintuitively, the full wave simulations (discrete green symbols) reveal that the transmission level increases with the incidence angle. Furthermore, startlingly, there is sharp transmission peak near $\theta^{inc} = 80°$, corresponding to an anomalous light tunneling through the metal wire meshes. For comparison, we provide in the supplementary materials the transmission characteristic of the individual meshes *A* and *B* at the same oscillation frequency [15]. As expected, the individual metallic meshes block very effectively the incoming radiation.

We developed an analytical model to uncover the physical principles underlying the light tunneling anomaly. Following Refs. [9, 10, 12], for long wavelengths a 3D metallic network can be modeled as an effective medium with dielectric function:

$$\frac{\overline{\varepsilon}}{\varepsilon_0} = \varepsilon_t(\omega)\left(\mathbf{1} - \frac{\mathbf{k} \otimes \mathbf{k}}{k^2}\right) + \varepsilon_l(\omega,k)\frac{\mathbf{k} \otimes \mathbf{k}}{k^2}, \qquad (1)$$

where **1** is the identity dyadic, $\otimes$ denotes the tensor product, $\varepsilon_t(\omega)$ is the transverse permittivity and $\varepsilon_l(\omega,k)$ is the longitudinal permittivity. The dielectric function depends explicitly on the wave vector $\mathbf{k} = -i\nabla$, and hence the effective medium response is spatially dispersive. Evidently, the dielectric function has contributions from both networks *A* and *B*. Under the hypothesis that the two networks interact with one another as homogenized media, the two contributions can be additively combined [9, 12, 20]. This approximation is more accurate when the physical distance between the two networks is larger; in this study the distance between the networks is the largest possible ($a/2$). Then, from Ref. [10, Eq. (32)] it is possible to write (supposing without loss of generality that the host medium is air):

$$\varepsilon_t(\omega) = 1 + \sum_{i=A,B} \left( \frac{1}{\varepsilon_{m,i}-1} \frac{1}{f_{V,i}} - \frac{\omega^2}{c^2 k_{p,i}^2} \right)^{-1}, \qquad (2a)$$

$$\varepsilon_l(\omega,k) = 1 + \sum_{i=A,B} \left( \frac{k^2}{l_{0,i} k_{p,i}^2} + \frac{1}{\varepsilon_{m,i}-1} \frac{1}{f_{V,i}} - \frac{\omega^2}{c^2 k_{p,i}^2} \right)^{-1}. \qquad (2b)$$

In the above, $\varepsilon_{m,i} = \varepsilon_{m,i}(\omega)$ ($i=A,B$) is the relative permittivity of the metal for the $i$-th wire mesh, $f_{V,i} = \pi r_{w,i}^2 / a^2$ is the volume fraction of the wires, $k_{p,i} = \frac{2\pi}{a} \left[ \ln\left(\frac{a}{2\pi r_{w,i}}\right) + 0.5275 \right]^{-1/2}$ is the effective plasma wave number and $l_{0,i} = 3/\left(1 + 2k_{p,i}^2/\beta_{1,i}^2\right)$ is a dimensionless slow wave factor that determines the strength of the nonlocal response [9, 10]. The slow wave factor is typically on the order of $l_{0,i} \sim 2$, and is written in terms of a parameter $\beta_{1,i}$, whose definition can be found at Refs. [9, 10]. Note that a local response would require $l_{0,i} \to \infty$; the small value of $l_{0,i}$ highlights that the effective medium is strongly nonlocal.

The dispersion of the transverse (TEM) modes is determined by $k^2 = (\omega/c)^2 \varepsilon_t(\omega)$. The TEM modes are doubly degenerate and have electromagnetic fields perpendicular to the wave vector. In the particular case of PEC wires ($\varepsilon_{m,i} = -\infty$) the TEM modes only propagate above the plasma frequency $\omega_{p,ef} = c\sqrt{k_{p,A}^2 + k_{p,B}^2}$, which is thereby larger than the plasma frequencies of the individual wire networks ($\omega_{p,i} = ck_{p,i}$, $i=A,B$). This is consistent with the heuristic idea that the interlaced wire meshes block more effectively the radiation than each individual wire network on its own. On the other hand, the longitudinal waves have a vanishing magnetic field and electric field directed along the wave vector. The dispersion of the longitudinal modes is determined by $\varepsilon_l(\omega,k) = 0$, and, provided the wires associated with the two networks are different (e.g., if the wires have different radii), it can be reduced to a quadratic polynomial equation in the variable $k^2$. This means that the interlaced wire medium supports *two* longitudinal modes. Figure 1b shows that on the overall there is a good agreement between the analytical model (solid lines) and the full wave numerical results (discrete points). Furthermore, consistent with the numerically simulated photonic band structure and with Ref. [13], one of the longitudinal modes predicted by our analytical model has no cut-off and propagates for arbitrarily low frequencies.

We build on ideas from our previous works [10, 12, 20, 21] to characterize with the analytical model the scattering of an incident plane wave by a finite thickness metamaterial slab. The total magnetic field ($\mathbf{H} = H_x \hat{\mathbf{x}}$) can be expanded in plane waves in all regions of space as follows (see the geometry in the inset of Fig. 2a):

$$H_x = H_0^{inc} e^{ik_y y} \times \begin{cases} e^{ik_z^0 z} + Re^{-ik_z^0 z}, & z \leq 0 \\ A_T^+ e^{ik_z^{(T)} z} + A_T^- e^{-ik_z^{(T)} z}, & 0 \leq z \leq L, \\ Te^{ik_z^0 (z-L)}, & z \geq L \end{cases} \qquad (3)$$

where $H_0^{inc}$ is the complex amplitude of the incident wave, $R$ and $T$ are the reflection and transmission coefficients, $A_T^\pm$ are the unknown amplitudes of the transverse mode in the wire medium, and $k_z^{(T)} = \sqrt{(\omega/c)^2 \varepsilon_t(\omega) - k_y^2}$ is the propagation constant along $z$ of the transverse wave in the wire medium. The incident wave vector is $\mathbf{k}_0^+ = k_y\hat{\mathbf{y}} + k_z^0\hat{\mathbf{z}}$ with $k_y = (\omega/c)\sin\theta^{inc}$ and $k_z^0 = \sqrt{(\omega/c)^2 - k_y^2}$. Note that the longitudinal modes have a vanishing magnetic field, and hence **H** depends only on the transverse mode in the interlaced wire medium.

In contrast, the electric field depends on both the transverse and longitudinal fields inside the wire medium and is given by,

$$\mathbf{E} = \frac{H_0^{inc}}{\omega\varepsilon_0}e^{ik_y y} \times \begin{cases} \hat{\mathbf{x}}\times\mathbf{k}_0^+ e^{ik_z^0 z} + \hat{\mathbf{x}}\times\mathbf{k}_0^- R e^{-ik_z^0 z}, & z \leq 0 \\ \sum_{\pm}\left(B_{L,1}^{\pm}\mathbf{k}_{L,1}^{\pm}e^{\pm ik_z^{(L,1)}z} + B_{L,2}^{\pm}\mathbf{k}_{L,2}^{\pm}e^{\pm ik_z^{(L,2)}z} + \frac{1}{\varepsilon_t}\hat{\mathbf{x}}\times\mathbf{k}_T^{\pm}A_T^{\pm}e^{\pm ik_z^{(T)}z}\right), & 0 \leq z \leq L, \quad (4) \\ \hat{\mathbf{x}}\times\mathbf{k}_0^+ T e^{ik_z^0(z-L)}, & z \geq L \end{cases}$$

where $k_z^{(L,j)}$ ($j$=1,2) are the $z$-propagation constants of the two longitudinal modes (which are found by solving $\varepsilon_l(\omega,k)=0$ with respect to $k_z$) and $B_{L,j}^{\pm}$ ($j$=1,2) are the unknown complex amplitudes of the longitudinal modes for the $\pm$ directions of propagation with respect to the $z$-axis. Furthermore, we defined $\mathbf{k}_0^{\pm} = k_y\hat{\mathbf{y}} \pm k_z^0\hat{\mathbf{z}}$, $\mathbf{k}_T^{\pm} = k_y\hat{\mathbf{y}} \pm k_z^{(T)}\hat{\mathbf{z}}$ and $\mathbf{k}_{L,j}^{\pm} = k_y\hat{\mathbf{y}} \pm k_z^{(L,j)}\hat{\mathbf{z}}$. Note that the transverse part of the electric field can be obtained from the magnetic field using $\mathbf{E}_T = \frac{1}{-i\omega\varepsilon_0\varepsilon_t}\nabla\times\mathbf{H}$.

The unknown coefficients ($R$, $T$, $A_T^{\pm}$, $B_{L,j}^{\pm}$) are found by imposing suitable boundary conditions at the $z=0$ and $z=L$ interfaces. As usual, we require the continuity of the tangential electromagnetic fields ($H_x$ and $E_y$) at the interfaces. However, due to the

-7-

nonlocal response of the interlaced wire medium, these Maxwellian boundary conditions are insufficient to determine all the unknowns and therefore one needs to enforce additional boundary conditions (ABCs) [10, 12, 21-24]. The standard (Pekar) ABC for a nonlocal plasma imposes that the normal component of the polarization current due to the drift charges ($-i\omega P_z$) vanishes at the interfaces [10, 25-27]. As shown in our previous works [18, 20, 21], this ABC can be extended to different types of wire media with multiple disconnected components. The idea is that the currents flowing in the individual meshes (here *A* and *B*) are independent, and therefore it is required that the contribution of each mesh to the polarization vector vanishes *separately* at the interfaces:

$$P_{z,i} = 0, \qquad (i=A,B) \qquad \text{at } z=0 \text{ and } z=L. \qquad (5)$$

The polarization vector associated with the *l*-th wire mesh (*l=A,B*) is $\mathbf{P}_l = (\bar{\varepsilon}_l - \varepsilon_0 \mathbf{1}) \cdot \mathbf{E}$, $\bar{\varepsilon}_l = \bar{\varepsilon}_l(\omega, -i\nabla)$ being the dielectric function of the *l*-th wire mesh alone. The dielectric function $\bar{\varepsilon}_l$ is defined as in Eqs. (1)-(2) except that the summations in Eq. (2) are restricted to the index $i = l$. Straightforward calculations show that $P_{z,i}$ can be explicitly written as follows:

$$P_{z,i} = \frac{H_0^{inc}}{\omega} e^{ik_y y} \sum_{\pm} \left( \pm k_z^{(L,1)} \left(\varepsilon_{l,i}^{(1)} - 1\right) B_{L,1}^\pm e^{\pm ik_z^{(L,1)}z} \pm k_z^{(L,2)} \left(\varepsilon_{l,i}^{(2)} - 1\right) B_{L,2}^\pm e^{\pm ik_z^{(L,2)}z} + \frac{\varepsilon_{t,i}-1}{\varepsilon_t} k_y A_T^\pm e^{\pm ik_z^{(T)}z} \right) \quad (6)$$

In the above, $\varepsilon_{t,i}(\omega) = 1 + \left( \frac{1}{\varepsilon_{m,i}-1} \frac{1}{f_{V,i}} - \frac{\omega^2}{c^2 k_{p,i}^2} \right)^{-1}$ is the transverse permittivity of the *i*-th mesh and $\varepsilon_{l,i}^{(j)} = 1 + \left( \frac{|\mathbf{k}_{L,j}|^2}{l_{0,i} k_{p,i}^2} + \frac{1}{\varepsilon_{m,i}-1} \frac{1}{f_{V,i}} - \frac{\omega^2}{c^2 k_{p,i}^2} \right)^{-1}$ is the corresponding longitudinal permittivity ($i=A,B$, $j=1,2$). In summary, the transmission and reflection coefficients can be found with the effective medium formalism by imposing the continuity of the tangential fields and two ABCs [Eq. (5)] per interface.

Using the outlined formalism, we calculated the transmission coefficient as a function of the incidence angle for the example discussed earlier. As shown in Fig. 2a, there is a truly remarkable agreement between the analytical model (solid line) and the CST full wave simulations. Figure 2b shows the transmission coefficient as function of the normalized thickness ($L/a$) for $\omega a/c = 1.32$. The analytical results (solid line) follow closely the full wave simulations (discrete symbols), which further validates our effective medium model and confirms the anomalous light tunneling. There are multiple transmission resonances, even for thicknesses as large as $L/a \sim 10$. The position of the transmission peaks is accurately predicted by the analytical model.

Figure 3a shows a density plot of the transmission coefficient as a function of $L$ and $\theta^{inc}$ for $\omega a/c = 1.32$. The density plot exhibits multiple sharp bright lines. The transmission is typically negligible for small angles; in fact, for normal incidence the propagating longitudinal mode cannot be excited. The transmission level becomes stronger for large incidence angles (grazing incidence) and for specific values of the thickness.

To understand the origin of the sharp lines, we investigated the Fabry-Pérot (FP) condition for the propagating longitudinal mode ($k_z^{(L,1)}L = n\pi$, $n = 1,2,3,...$). Figure 3b represents the combination of parameters $L$ and $\theta^{inc}$ required to have a FP resonance of order $n$. Note that for a fixed frequency $k_z^{(L,1)}$ is a function of the incidence angle. As seen, there is an obvious coincidence between the FP-resonance lines in Fig. 3b and the sharp bright lines in Fig. 3a. This clearly proves that the tunneling anomaly is due to a FP resonance of the propagating longitudinal mode. In the supplementary materials [15], we further examine the tunneling condition for a fixed slab thickness and varying $\omega$ and $\theta^{inc}$.

It is illuminating to see the profiles of the polarization currents in each wire mesh obtained with the effective medium model. Figure 4 depicts the profiles of $P_{z,A}$ and $P_{z,B}$ [see Eq. (6)] for the first two peaks of the transmission coefficient in Fig. 2b. As expected, the field profiles are consistent with FP resonances of $1^{st}$ and $2^{nd}$ order. However, the most remarkable thing in Fig. 4 is that it reveals that at the transmission resonances the polarization vectors $P_{z,A}$ and $P_{z,B}$, and thereby the currents in the $A$ and $B$ meshes, have nearly identical amplitudes and are phase-shifted by 180º ($\phi_A - \phi_B = 180º$). Thus, the tunneling anomaly is made possible by a *destructive interference* of the fields radiated by the two wire meshes. Indeed, for long wavelengths the scattering by the interlaced wire medium is typically dominated by a strongly radiative dipolar mode due to the in-phase interference of the fields scattered by the two wire meshes. Crucially, because of the structural asymmetry of interlaced wire medium ($r_{w,A} \neq r_{w,B}$) the currents excited in meshes $A$ and $B$ may be different. This enables the formation of a narrow anti-bonding mode, such that the current in one of the meshes flips sign for a narrow range of incident angles leading to a sub-radiant regime. Thus, the physical origin of the tunneling anomaly is a Fano-type resonance [20, 28, 29, 30] that enables the *cancellation of the scattering* by the two subcomponents of the interlaced wire medium. Remarkably, in our system the two scatterers (i.e. the wire meshes) that create the Fano resonance are infinitely extended in space.

To study the impact of metallic loss on the described tunneling anomaly, it is supposed next that the wires have a finite conductivity $\sigma_i$ (*i*=A,B) and that the lattice constant is $a = 1\,\text{mm}$. Thus, the metal complex permittivity is taken equal to $\varepsilon_{m,i} = 1 - \sigma_i / (-i\omega\varepsilon_0)$. Figures 2c and 2d show the amplitude of the transmission coefficient for different values of the conductivity $\sigma_B$ of the wires in mesh B. As before, the wires in mesh A are assumed PEC. For the copper case ($\sigma_{B,Cu} = 5.96 \times 10^7 \,\text{S}\,\text{m}^{-1}$) the loss effects are negligible, which is expected because the metal skin depth at the considered frequency ($\delta_B = \sqrt{2/(\omega\mu_0\sigma_B)} = 0.26\,\mu\text{m}$) is much smaller than the corresponding wire radius $r_{w,B} = 0.05\,\text{mm}$. For the conductivity $\sigma_B = 10^5\,\text{S}\,\text{m}^{-1}$ ($\delta_B = 6.34\,\mu\text{m}$), the transmission peaks are evidently damped because a significant fraction of the fields energy can penetrate inside the lossy metal.

In summary, we theoretically predicted the counterintuitive effect of light tunneling due to the destructive interference of the scattering by two interlaced 3D wire meshes. The effect is rooted in a sub-radiant Fano-type resonance and was explained with an effective medium theory that models the interlaced wire medium as a spatially dispersive continuum. It was shown that the transmission peaks correspond to FP resonances of a longitudinal propagating mode supported by the metamaterial. It is underlined that the longitudinal mode can be excited by a (slowly varying in space) plane wave only due to the structural asymmetry of the wire meshes. The structural asymmetry is mandatory in order that the currents in the two wire meshes can be 180º out of phase. Since the transmission peaks are very sensitive to changes of the incidence angle, frequency and of other structural parameters, the interlaced wire medium may find interesting applications in sensing and angular filtering.

**Acknowledgement**: This work is supported in part by Fundação para a Ciência e a Tecnologia with the grants PTDC/EEITEL/4543/2014 and UID/EEA/50008/2013. Hafssaa Latioui acknowledges financial support by the Erasmus Mundus programme of the European Union - Al-Idrisi II.

**Figures**

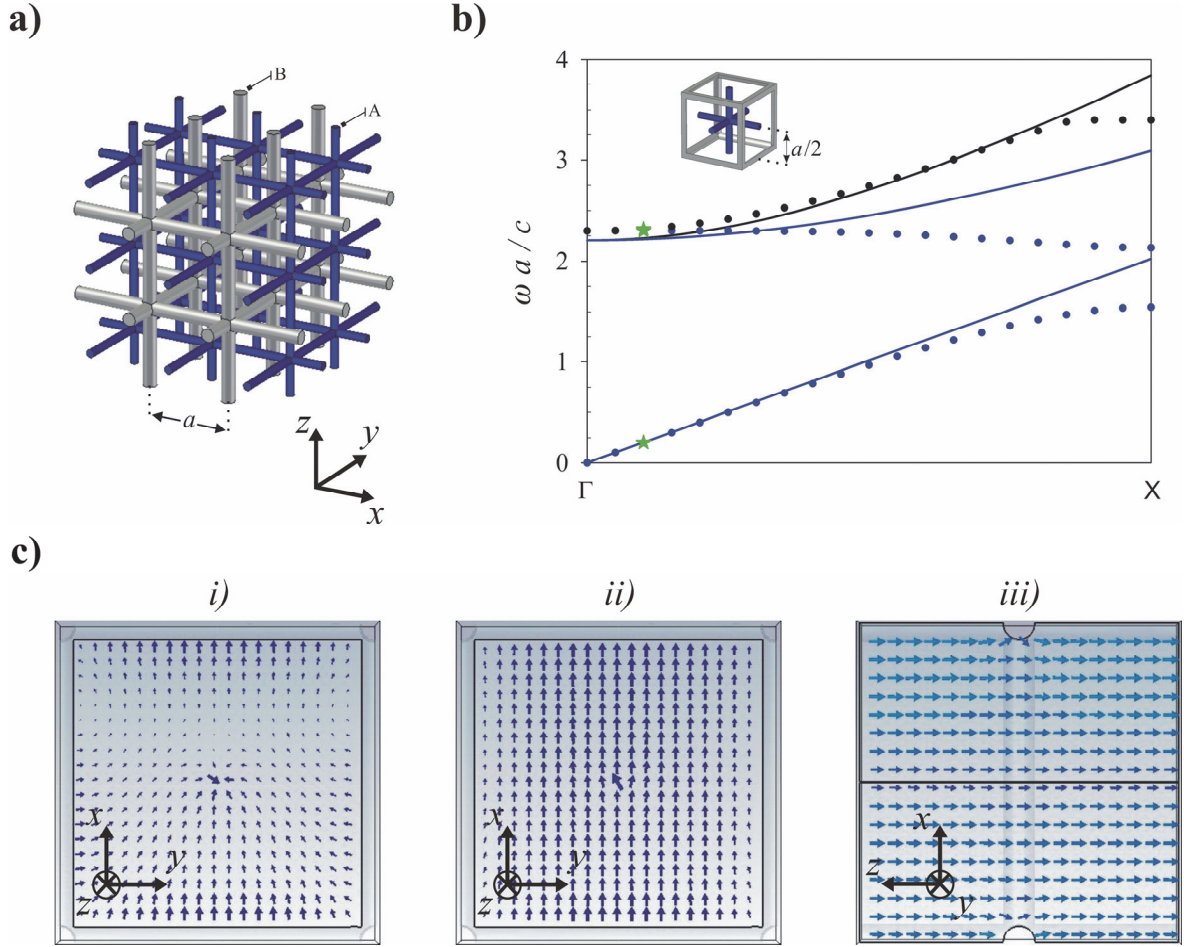

**Fig. 1. a)** Geometry of the interlaced wire meshes. The wires of each network are spaced by a distance $a$ along the coordinate axes. The distance between the two non-connected networks is $a/2$. **b)** Band diagram of the electromagnetic modes along the direction $\Gamma X$. Solid lines: analytical model; Discrete symbols: full wave simulations. The inset shows the cubic unit cell of the structure. The wires are PEC and are embedded in a dielectric with permittivity $\varepsilon_h = 1$; the radii of the wires are: $r_{w,A} = 0.001a$ and $r_{w,B} = 0.05a$. **c)** Electric field profile for *i)* Low-frequency longitudinal mode at $\omega a/c = 0.20$, *ii)* High-frequency longitudinal mode at $\omega a/c = 2.30$, *iii)* One of the degenerate TEM modes at $\omega a/c = 2.32$. The modes are marked with green stars in panel b).

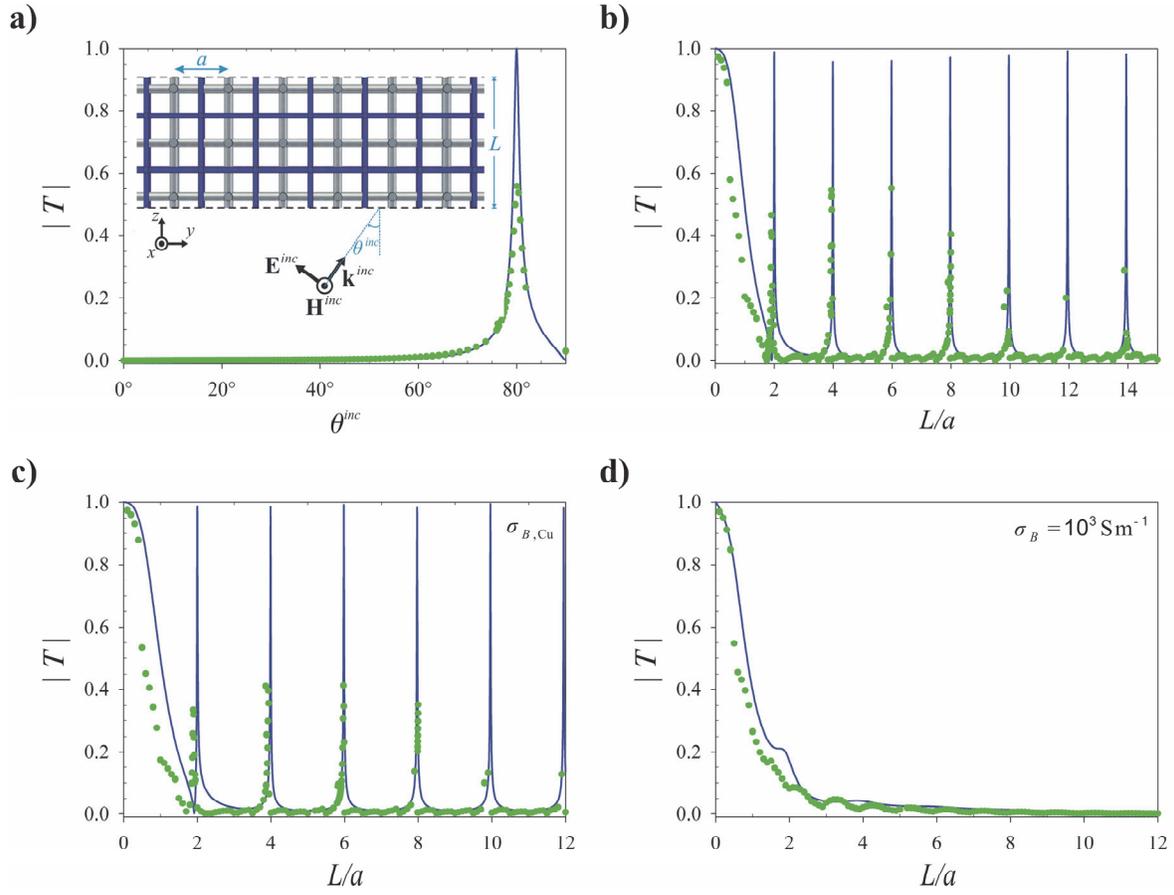

**Fig. 2**. **a)** Amplitude of the transmission coefficient as function of the incidence angle for the normalized frequency $\omega a / c = 1.32$ and normalized thickness $L/a \approx 6$. The remaining structural parameters are as in Fig. 1. The inset shows the geometry of the problem. **b)** Amplitude of the transmission coefficient as a function of the normalized thickness for the fixed frequency $\omega a / c = 1.32$ and incidence angle 80º. **c)** and **d)** Similar to b) but for wires with different conductivity values and $a = 1\,\text{mm}$. In all the panels the solid lines represent the analytical results, and the discrete symbols the full wave simulations results.

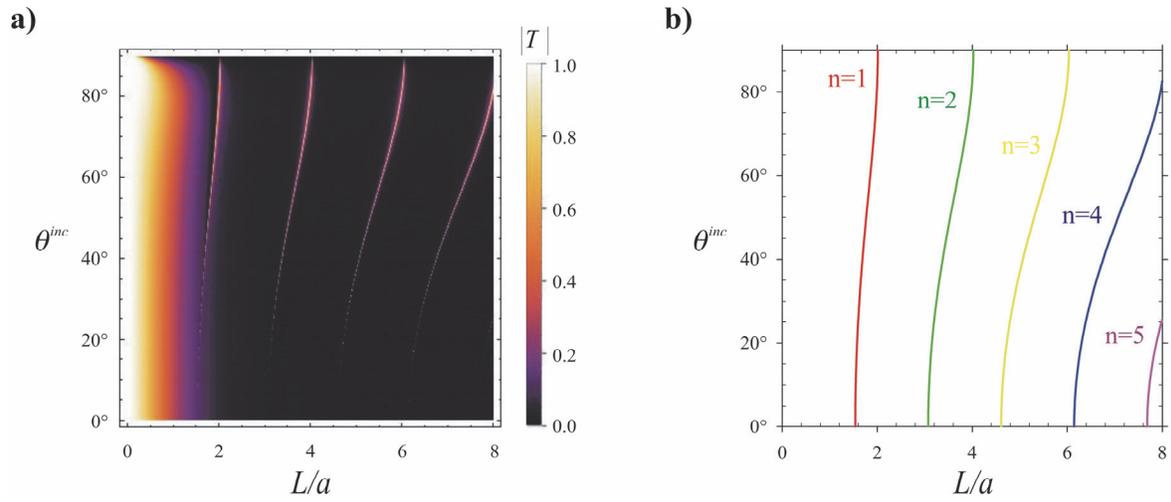

**Fig. 3**. **a)** Density plot of the transmission coefficient amplitude as a function of the normalized thickness $L/a$ and of the incidence angle $\theta^{inc}$ at the fixed frequency of $\omega a/c = 1.32$. **b)** Incidence angle $\theta^{inc}$ as a function of $L/a$ for the $n$-th ($n$=1,2,…) Fabry-Pérot resonance of the propagating longitudinal mode at $\omega a/c = 1.32$.

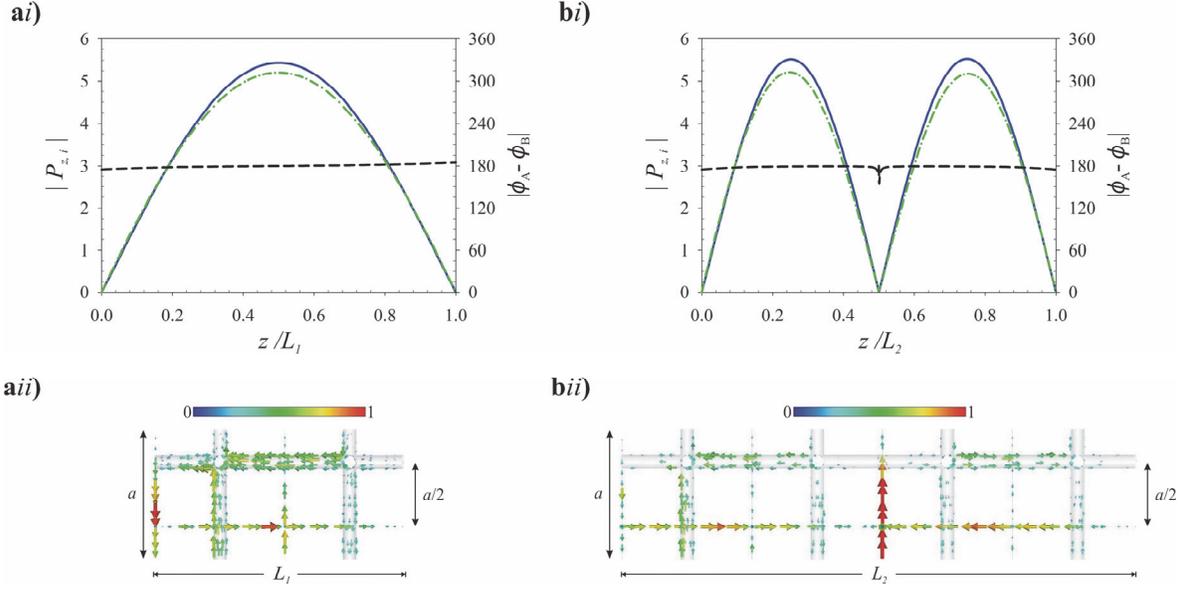

**Fig. 4.** (**a***i*, **b***i*) Phase difference (black dashed lines) and normalized amplitude (blue and green lines) of the *z*-component of the macroscopic polarization $P_{z,A}$ and $P_{z,B}$ [Eq. (6)] associated with the wire meshes *A* and *B* for **a***i*) the 1$^{st}$ FP resonance ( $L/a \approx 2$ ) and **b***i*) the 2$^{nd}$ FP resonance ( $L/a \approx 4$ ). The solid blue (dot-dashed green) lines represent the contribution from the wire mesh *A* (wire mesh *B*). The normalized frequency is $\omega a / c = 1.32$ and the incidence angle is 80º. The remaining simulation parameters are as in Fig. 2b). (**a***ii*, **b***ii*) Full wave simulation results of the microscopic current density for the scenarios **a***i*) (1$^{st}$ FP resonance) and **b***i*) (2$^{nd}$ FP resonance), respectively.

# Supplemental Material for the Manuscript

# "Light tunneling anomaly in interlaced metallic wire meshes"

Hafssaa Latioui[1], Mário G. Silveirinha[1,2*]

[1]*Instituto de Telecomunicações and Department of Electrical Engineering, University of Coimbra, 3030-290 Coimbra, Portugal*

[2]*University of Lisbon, Instituto Superior Técnico, Avenida Rovisco Pais, 1, 1049-001 Lisboa, Portugal*

In this document we present *(i)* the photonic band structure of the individual wire meshes, *(ii)* the transmission characteristic of the individual wire meshes, and *(iii)* a study of the tunneling condition for a varying frequency and incidence angle.

## A. Band structure and transmission characteristics of the individual wire meshes

We consider two independent 3D metallic wire meshes "A" and "B". The metal is assumed to be a perfect electric conductor (PEC) and the host material is air. As in the main text, the wire radius is $r_{w,A} = 0.001a$ for the wire mesh A and $r_{w,B} = 0.05a$ for the wire mesh B, where *a* is the lattice period. Figure S1 shows the photonic band structure calculated with CST Microwave Studio (discrete symbols) and with the analytical model (solid lines) of Refs. [S1, S2]. As seen, the individual wire meshes do not support light states in the long wavelength regime [S1, S2].

---

[*] To whom correspondence should be addressed: E-mail: mario.silveirinha@co.it.pt

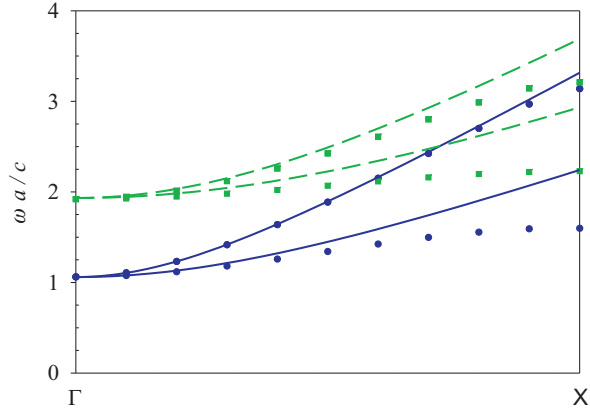

**Fig. S1.** Band diagram of the electromagnetic modes along the $\Gamma X$ direction for each individual wire mesh. Solid lines: analytical model; Discrete symbols: full wave simulations [S3]. The wires are PEC and are embedded in air. The lattice period is $a$. Blue color: wire mesh A with $r_{w,A} = 0.001a$. Green color: wire mesh B with $r_{w,B} = 0.05a$.

Next, we consider the case where a metamaterial slab is excited by an incident plane wave polarized as shown in the inset of Fig. S2, with the incidence angle $80°$. The oscillation frequency is fixed as $\omega a / c = 1.32$. Figure S2 depicts the amplitude of the transmission coefficient as function of the normalized thickness $L/a$ for each of the individual wire meshes. As seen, for a sufficiently thick metamaterial slab the transmission level is rather weak. Note that in this example $\omega$ is less than the effective plasma frequency for mesh B ($\omega_{p,ef}^B \approx 1.93c/a$), but greater than the effective plasma frequency for mesh A ($\omega_{p,ef}^A \approx 1.06c/a$). Even though the mesh A supports propagating modes, the transmission level is rather weak because of the large incidence angle. Indeed, a standard metallic mesh typically blocks very effectively the incoming radiation for grazing incidence, even when the frequency is larger than the effective plasma frequency.

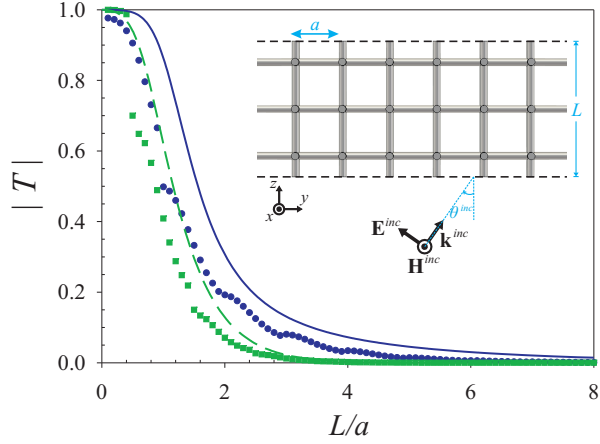

**Fig. S2.** Amplitude of the transmission coefficient as function of the normalized thickness for the fixed frequency $\omega a / c = 1.32$ and incidence angle $80°$. Blue color: wire mesh A with $r_{w,A} = 0.001a$. Green color: wire mesh B with $r_{w,B} = 0.05a$. Solid lines: analytical model [S2]; Discrete symbols: full wave simulations [S3]. The inset shows the geometry of the problem.

## *B. Light tunneling condition*

Next, we investigate the light tunneling condition for a fixed interlaced wire medium slab thickness ($L / a \approx 6$) and for a varying frequency and incidence angle (see the inset of Fig. 2a of the main text for the problem geometry).

Figure S3a shows a density plot of the transmission coefficient amplitude as a function of $\omega a / c$ and $\theta^{\text{inc}}$. Similar to Fig. 3 of the main text, the density plot is characterized by multiple sharp lines, which are especially "bright" for large incidence angles. Figure S3b identifies the combinations of parameters $\omega a / c$ and $\theta^{\text{inc}}$ associated with a Fabry-Pérot (FP) resonance of order *n* of the propagating longitudinal mode ($k_z^{(L,1)} L = n\pi$, $n = 1, 2, 3, ...$). As seen, there is an exact matching between the FP-resonance lines in Fig. S3b and the sharp bright lines in Fig. S3a, confirming that the tunneling anomaly is due to a FP resonance of the propagating longitudinal mode. Moreover, the results of Fig. S3 reveal that the tunneling anomaly can in principle occur for extremely long wavelengths ($\omega a / c \ll 1$). It is possible to further red-shift the transmission resonance by increasing the value of $L/a$ (not shown). For grazing incidence and for the considered wire radii, the first resonance occurs for $\omega L/c \sim 2.6$, approximately independent of the value of $L/a$.

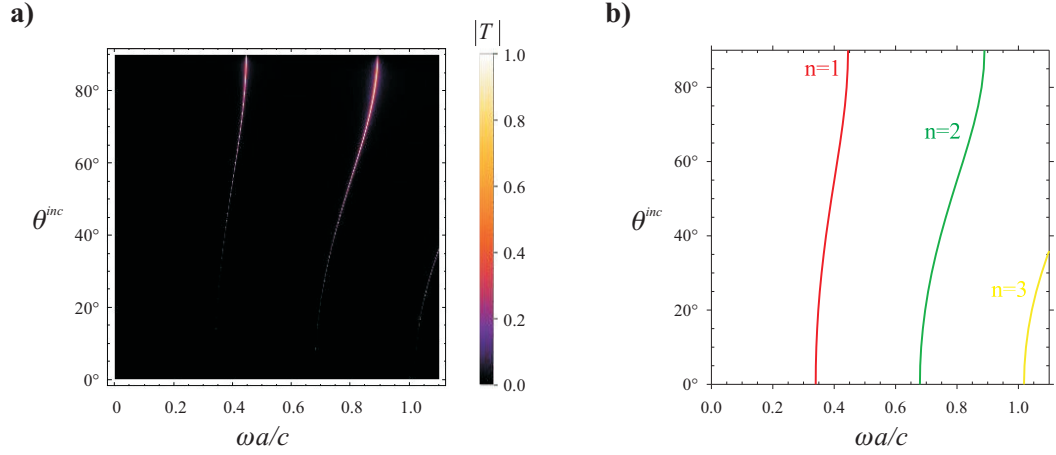

**Fig. S3. a)** Density plot of the transmission coefficient amplitude as a function of the normalized frequency $\omega a/c$ and of the incidence angle $\theta^{inc}$ for the fixed slab thickness $L/a \approx 6$. **b)** Incidence angle $\theta^{inc}$ as a function of the normalized frequency $\omega a/c$ for the $n^{th}$ ($n$=1,2,…) Fabry-Pérot resonance of the propagating longitudinal mode and $L/a \approx 6$.